\documentclass[preprint, authoryear, 3p]{elsarticle} 

\usepackage[utf8]{inputenc}

\usepackage{amsmath}
\usepackage{amsthm}
\usepackage{amssymb}
\usepackage[table]{xcolor}
\usepackage{mathtools}
\usepackage{bm}
\usepackage{setspace}
\usepackage{subcaption}
\usepackage{natbib}
\usepackage{hyperref}
\usepackage{float}

\usepackage[title]{appendix}

\newcommand{\E}[0]{\ensuremath{\mathtt{E}}-}
\newcommand{\kl}{\ensuremath{\text{\sc kl}}}

\newcolumntype{C}[1]{>{\centering\arraybackslash}m{#1}}
\newcolumntype{R}[1]{>{\raggedleft\arraybackslash}m{#1}}
\definecolor{mygray}{gray}{0.9}

\newtheorem{theorem}{Theorem}

\theoremstyle{plain}

\newcommand{\vecspace}{\ensuremath{\vec{\Theta}}}
\newcommand{\vecpar}{\ensuremath{\vec{\theta}}}

\newcommand{\commentout}[1]{}

\newcommand{\cH}{\ensuremath{\mathcal{H}}}

\newcommand{\cW}{\ensuremath{\mathcal{W}}}
\newcommand{\cY}{\ensuremath{\mathcal{Y}}}
\newcommand{\naturals}{\ensuremath{\mathbf N}}
\newcommand{\reals}{\ensuremath{\mathbf R}}

\renewcommand{\sp}{\ensuremath{\theta_a,\theta_b}}
\newcommand{\ta}{\ensuremath{\theta^*_a}}
\newcommand{\tb}{\ensuremath{\theta^*_b}}

\newtheorem{examplehidden}{Example}

\title{Exact Anytime-valid Confidence Intervals for Contingency Tables and Beyond}
\author[1,2]{Rosanne J. Turner\corref{cor1}}
\author[1,3]{Peter D. Gr\"unwald}
\address[1]{CWI, Amsterdam, part of NWO-I, Netherlands}
\address[2]{University Medical Center Utrecht, Brain Center, Netherlands}
\address[3]{Leiden University, Department of Mathematics, Netherlands}
\cortext[cor1]{Corresponding author, Rosanne.Turner@cwi.nl, National Research Institute for mathematics and computer science in the Netherlands (CWI),
Science Park 123, 1098 XG Amsterdam,
The Netherlands. Declarations of interest: none.}
\date{March 2022}
\begin{document}
\begin{abstract} 
E-variables are tools for retaining type-I error guarantee  with optional stopping. We extend E-variables for sequential two-sample tests to general null hypotheses and anytime-valid confidence sequences. We provide implementations for estimating risk difference, relative risk and odds-ratios in contingency tables.
\end{abstract}
\begin{keyword}
    confidence sequences \sep contingency tables \sep anytime-valid \sep effect size
\end{keyword}
\maketitle
\doublespacing
\section{Introduction}\label{sec:introduction}
We consider a setting where we collect samples from two distinct groups, denoted $a$ and $b$. 
In both groups, data come in sequentially and are i.i.d. We thus have two data streams, $Y_{1,a}, Y_{2,a}, \ldots$ i.i.d. $\sim P_{\theta_a}$ and $Y_{1,b}, Y_{2,b},\ldots $ i.i.d. $\sim P_{\theta_b}$ where we assume that $\theta_a, \theta_b \in \Theta$, $\{P_{\theta} : \theta \in \Theta\}$ representing some parameterized underlying family of distributions, all assumed  to have a probability density or mass function denoted by  $p_{\theta}$ on some outcome space $\cY$.

\E variables \citep{grunwald2019safe, VovkW21} are a tool for constructing tests that keep their Type-I error control under optional stopping and continuation.  Previously, \cite{turner2021safe} developed \E variables for testing equality of both data streams, i.e. with null hypothesis $\vecspace_0 := \{(\theta_a,\theta_b) \in \Theta^2 : \theta_a = \theta_b \}$. Here we first generalize these \E variables to more general null hypotheses in which we may have $\theta_a \neq \theta_b$. We then use these generalized \E variables to construct  {\em anytime-valid\/} confidence sequences; these provide confidence sets that remain valid under optional stopping 
\citep{darling1967confidence,howard2018uniform}. 

As in \citep{turner2021safe}, we first design \E variables for a  {\em single block\/} of data $(Y_a^{n_a},Y_b^{n_b})$, where a  block is a set of data  consisting of $n_a$ outcomes $Y_a^{n_a} = (Y_{a,1}, \ldots, Y_{a,n_a})$ in group $a$ and $n_b$ outcomes $Y_b^{n_b} = (Y_{b,1}, \ldots, Y_{b,n_b})$ in group $b$, for some pre-specified $n_a$ and $n_b$. An {\em \E variable\/} is then, by definition, any   nonnegative random variable $S = s'(Y_a^{n_a},Y_b^{n_b})$ such that 
\begin{equation}
   \label{eq:thirdc}
\sup_{(\theta_a,\theta_b) \in \vecspace_0} {\bf E}_{Y_a^{n_a} \sim  P_{\theta_a}, Y_b^{n_b} \sim P_{\theta_b}}\left[s'(Y^{n_a}_a, Y^{n_b}_b) \right] \leq 1.
\end{equation}
\cite{turner2021safe} first defined such an \E variable for  $\vecspace_0 =  \{ (\sp) \in \Theta^2: \theta_a= \theta_b\}$ so that it would tend to have high power
against a given simple alternative $\vecspace_1 = \{(\theta^*_a,\theta^*_b)\}$.
Their \E variable is of the following simple form (with $n=n_a+n_b$):
\begin{equation}\label{eq:simpler}
s'(Y^{n_a}_a,Y^{n_b}_b) = \frac{p_{\theta^*_a}(Y_a^{n_a})}{\prod_{i=1}^{n_a} (\frac{n_a}{n}
p_{\theta^*_a}(Y_{a,i}) + \frac{n_b}{n}
p_{\theta^*_b}(Y_{a,i}))
}
\cdot 
\frac{p_{\theta^*_b}(Y_b^{n_b})}{
\prod_{i=1}^{n_b}
(\frac{n_a}{n}
p_{\theta^*_a}(Y_{b,i}) + \frac{n_b}{n}
p_{\theta^*_b}(Y_{b,i}))
}.
\end{equation}
These \E variables can be extended to sequences of blocks $Y_{(1)}, Y_{(2)}, \ldots$ by multiplication, and can be extended to composite alternatives by sequentially learning $(\theta^*_a,\theta^*_b)$ from the data, for example via a Bayesian prior on $\vecspace_1$. The $n_a$ and $n_b$ used for the $j$-th block $Y_{(j)}$ are allowed to depend on past data, but they must be fixed before the first observation in block $j$ occurs. For simplicity, in this note we only consider the case with $n_a$ and $n_b$ that remain fixed throughout; extension to the general case is straightforward.

By a general property of \E variables, at each point in time, the running product of block \E variables observed so far is itself an \E variable, and the random process of the products is known as a {\em test martingale} \citep{grunwald2019safe, shafer2020language}. An \E variable-based test at level $\alpha$ is a test which, in combination with any stopping rule $\tau$, reports `reject' if and only if the product of \E values corresponding to all blocks that were observed at the stopping time and have already been completed, is larger than $1/\alpha$. Such a test has a type-I error probability bounded by $\alpha$ irrespective of the stopping time $\tau$ that was used; see the aforementioned references for much more detailed introductions and, for example \citep{henzi2021valid}, for a practical application. 

In case $\{P_{\theta} : \theta \in \Theta \}$ is convex, the \E variable (\ref{eq:simpler}) has the so-called GRO-({\em growth-rate-optimality\/}) property: it maximizes, over all \E variables (i.e. over all nonnegative random variables $S= s'(Y_a^{n_a},Y_b^{n_b})$ satisfying (\ref{eq:thirdc})) the logarithmic growth rate
\begin{equation}\label{eq:grow}
    {\bf E}_{Y_a^{n_a} \sim  P_{\theta^*_a}, Y_b^{n_b} \sim P_{\theta^*_b}}\left[
    \log S\right],
\end{equation}
which implies that, under $(\theta^*_a,\theta^*_b)$, the expected number of data points before the null can be rejected is minimized \citep{grunwald2019safe}.

Below, in Theorem~\ref{simpleSproportionsgenH0} in section \ref{sec:extensions}, which generalizes Theorem 1 in \cite{turner2021safe}, we 
extend (\ref{eq:simpler}) to the case of general null hypotheses, $\vecspace_0 \subset \Theta^2$, allowing for the case that the  elements of $\vecspace_0$ have two different components, and provide a condition under which it has the GRO property. From then onwards we focus on what we call `the $2 \times 2$ contingency table setting' in which both streams are Bernoulli, $\theta_j$ denoting the probability of $1$ in group $j$. For this case, Theorem~\ref{thm:convex} gives a simplified expression for the \E variable and shows that the GRO property holds if $\vecspace_0 \subset [0,1]^2$ is convex. 
Then we will extend this \E variable to deal with composite $\vecspace_1$ and use this to define anytime-valid confidence sequences. We illustrate these through simulations. All proofs are in Appendix \ref{app:proofs}.

\section{General Null Hypotheses}
\label{sec:extensions}
In this section, we first construct an \E variable for general null hypotheses that generalizes (\ref{eq:simpler}).  We then instantiate the new result to the $2 \times 2$ case. 
The following development and results require $\{P_{\theta} :\theta \in \Theta\}$ to be  `nondegenerate' in the sense that there exists $\theta\in \Theta$ such that for all $\theta' \in \Theta$, $D(P_{\theta}\| P_{\theta'})< \infty$. This mild condition holds, for example, for exponential families; we tacitly assume nondegeneracy from now on. 

Our goal is thus to define an \E variable for a block of $n= n_a + n_b$ data points with $n_g$ points in group $g$, $g \in \{a,b \}$. For notational convenience we define, for $\theta_a,\theta_b \in \Theta$,  $P_{\theta_a,\theta_b}$ as the joint distribution of $Y_a^{n_a} \sim P_{\theta_a}$ and $Y_b^{n_b} \sim P_{\theta_b}$, so that $p_{\theta_a,\theta_b}(y_a^{n_a},y_b^{n_b}) = \prod_{i=1}^{n_a} p_{\theta_a}(y_{a,i}) \prod_{i=1}^{n_b} p_{\theta_b}(y_{b,i})$ so that we can write the null hypothesis as $\cH_0 := \{P_{\theta_a,\theta_b} : (\theta_a,\theta_b) \in \vecspace_0 \}$.
Our strategy will be to first develop an \E-variable for a {\em modified\/} setting 
in which there is only a single outcome, falling with probability $n_a/n$ in group $a$ and $n_b/n$ in group $b$. 
To this end, for $\vecpar = (\theta_a,\theta_b)$, we define $p'_{\vecpar}(Y|a) := p_{\theta_a}(y), p'_{\vecpar}(Y|b) := p_{\theta_b}(y)$, all distributions with a $'$ refering to the modified setting with just one outcome. We let $\cW^{\circ}(\vecspace_0)$ be the set of all distributions on $\vecspace_0$ with finite support. For $W \in \cW^{\circ}(\vecspace_0)$, we define $p'_W(Y|g)= \int p'_{\vecpar}(Y|g) d W(\vecpar)$.
We set $p'_W(y^k |g) := \prod_{i=1}^k p'_W(y_i |g)$. We further define, for given alternative $\vecspace_1 = \{(\theta^*_a,\theta^*_b)\}$,  $p^{\circ}(\cdot|g)$, $g \in \{a,b\}$ to be, if it exists, the conditional probability density satisfying 
 \begin{equation}
     \label{eq:klmina}
     {\bf E}_{G \sim Q'} {\bf E}_{Y \sim P_{\theta^*_G}} \left[ - \log p^{\circ}(Y \mid G) \right] = 
\inf_{W \in \cW^{\circ}(\vecspace_0)} {\bf E}_{G \sim Q'} {\bf E}_{Y \sim P_{\theta^*_G}} \left[ - \log {p}'_W(Y \mid G) \right]
\end{equation}
with $Q'(G)$ the distribution for $G \in \{a,b\}$ with $Q'(G=a) = n_a/n$.
Clearly we can rephrase (\ref{eq:klmina}) equivalently as: 
\begin{equation}\label{eq:klminb}
D(Q'(G,Y) \| P^{\circ}(G,Y))
 =  \inf_{W \in \cW^{\circ}(\vecspace_0)}
  D(Q'(G,Y) \| P'_W(G,Y)),
\end{equation}
where $D$ is the KL divergence. Here we extended the conditional distributions $P'_W(Y|G)$ and $P^{\circ}(Y|G)$ (corresponding to densities $p'_W(Y|G)$ and $p^{\circ}(Y|G)$) to a joint distribution by setting $P'_W(G,Y) := Q'(G) P'_W(Y|G)$ (and similarly for $P^{\circ}$) and we extended $Q'(G,Y) :=
Q'(G) P_{\theta^*_G}(Y)$. 
We have now constructed a modified null hypothesis $\cH'_0 = \{P'_{\vecpar}(G,Y): \vecpar \in \vecspace_0\}$ of joint distributions for a single `group' outcome $G \in \{a,b \}$ and `data' outcome $Y \in \cY$. We let  $\bar{\cH}'_0 = \{ P_W(G,Y): W \in \cW^{\circ}(\vecspace_0) \}$ be the convex hull of $\cH'_0$.

The $p^{\circ}$ satisfying (\ref{eq:klminb}) is commonly called  the {\em reverse information projection\/}  of $Q'$ onto $\bar{\cH}'_0$. 
\cite{Li99} shows that $p^{\circ}$ always exists under our nondegeneracy condition, though in some cases it may represent a sub-distribution (integrating to strictly less than one); see  
 \cite[Theorem 1]{grunwald2019safe} (re-stated for convenience in the supplementary material) who, building on Li's work, established a general relation between reverse information projection and \E-variables. Part 1 of that theorem
 establishes that if  the minimum in (\ref{eq:klmina}) (or (\ref{eq:klminb})) is achieved by some $W^{\circ} \in \cW^{\circ}$ then $p^{\circ}(\cdot |\cdot) = p'_{W^{\circ}}(\cdot |\cdot)$ and, with $\vecpar^* = (\theta_a,\theta_b)$, for all $\vecpar \in \vecspace_0$,  
\begin{equation}\label{eq:maandag}
    {\bf E}_{G \sim Q'} {\bf E}_{Y \sim P'_{\vecpar}|G} \left[ 
    \frac{p'_{\vecpar^*}(Y|G)}{
p^{\circ}(Y|G)}
    \right] 
    = {\bf E}_{G \sim Q'} {\bf E}_{Y \sim P'_{\vecpar}|G} \left[ 
    \frac{p'_{\vecpar^*}(G,Y)}{
p^{\circ}(G,Y)}
    \right] 
    \leq 1. 
\end{equation}
This expresses that $p'_{\vecpar^*}(Y|G)/p^{\circ}(Y|G)$ is an \E variable for our modified problem, in which within a single block we observe a single outcome in group $g$, with $g$ chosen with probability $n_g/n$. 
If we were to interpret the  \E-variable of the  modified problem as in (\ref{eq:maandag}) as a likelihood ratio for a single outcome, its corresponding likelihood ratio for a single block of data in our original problem with $n_g$ outcomes in group $g$ would be:
\begin{equation}\label{eq:simpleevargenH0}
s(y^{n_a}_a, y^{n_b}_b; n_a, n_b, (\theta^*_a,\theta^*_b); \vecspace_0)
\coloneqq  \frac{p'_{(\theta^*_a,\theta^*_b)}( y^{n_a}_a|a)
p'_{(\theta^*_a,\theta^*_b)}( y^{n_b}_b|b)}{
p^{\circ}(y^{n_a}_a |a)
p^{\circ}(y^{n_b}_b |b)} =  \frac{p_{\theta_{a}^*}( y^{n_a}_a) p_{\theta_b^*}(y^{n_b}_b)}{
p^{\circ}(y^{n_a}_a |a)
p^{\circ}(y^{n_b}_b |b)}.
 \end{equation}
The following theorem expresses that this `extension' of the \E variable in the modified problem gives us an \E variable in our original problem: 
\begin{theorem}
\label{simpleSproportionsgenH0}
$S_{[n_a,n_b, \theta^*_a,\theta^*_b; \vecspace_0]} :=  s(Y^{n_a}_a, Y^{n_b}_b; n_a,n_b, (\theta^*_a,\theta^*_b); \vecspace_0)$ as in (\ref{eq:simpleevargenH0}) is an $E$-variable, i.e. with $s'(\cdot)= s( \cdot ; n_a,n_b, (\theta^*_a,\theta^*_b);\vecspace_0)$, we have (\ref{eq:thirdc}). 
Moreover, if $\cH'_0 = \{P'_{\vecpar}: \vecpar \in \vecspace_0 \}$ (the null hypothesis for the {\em modified\/} problem) is a convex set of distributions  and $\cY$ is finite (so that $\cH'_0 = \bar{\cH}'_0$) and furthermore $\cH'_0$ is compact in the weak topology, then (a) $p^{\circ}(\cdot |\cdot) = p'_{\vecpar}(\cdot |\cdot)$ for some $\vecpar \in \vecspace_0$ and (b)  $S_{[n_a,n_b, \theta^*_a,\theta^*_b; \vecspace_0]}$ is the $(\theta^*_a,\theta^*_b)$-GRO \E variable for the {\em original\/} problem, maximizing (\ref{eq:grow}) among all \E variables.
\end{theorem}
In the case that $\cH'_0$ is not convex and compact, we do not have a simple expression for $p^{\circ}$ in general, and we may have to find it numerically by minimizing (\ref{eq:klmina}). In the $2 \times 2$ table (Bernoulli $\Theta$) case though, there are interesting $\cH_0$  for which the corresponding $\cH'_0$ is convex, and we shall now see that this leads to major simplifications.  
\subsection{General Convex $\vecspace_0$ for the $2 \times 2$ contingency table}\label{sec:extb}
\begin{figure}[ht]
     \centering
     \begin{subfigure}[b]{4cm}
         \centering
         \includegraphics[width=\textwidth]{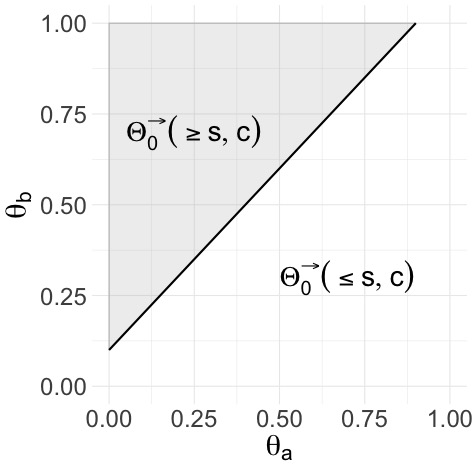}
         \caption{linear boundary}
     \end{subfigure}
     \hspace{1cm}
     \begin{subfigure}[b]{4cm}
         \centering
         \includegraphics[width=\textwidth]{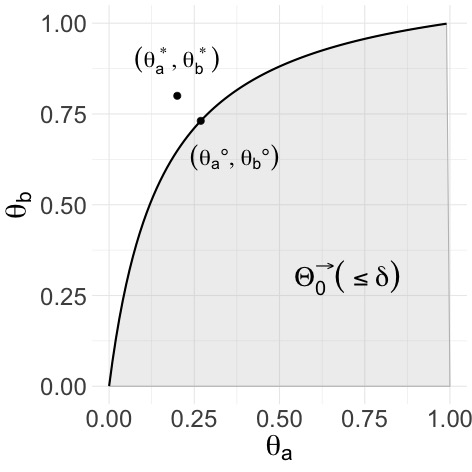}
         \caption{log odds ratio boundary}
     \end{subfigure}
        \caption{Examples of null hypothesis parameter spaces for two types of boundaries.}
        \label{fig:parameterSpaceExamples}
\end{figure}

In this subsection and the next, $\{ P_{\theta_a,\theta_b} \}$ refers to the $2 \times 2$ model again, with $\cY = \{0,1\}$ and $\theta$ denoting the probability of $1$. We now let $\vecspace_0$ be any closed convex subset of $[0,1]^2$ that contains a point in the interior of $[0,1]^2$. Again, note that the corresponding $\cH_0 = \{ P_{\vecpar}: \vecpar \in \vecspace_0 \}$ need not be convex; still, $\cH'_0$, the null hypothesis for the modified problem as defined above, must be convex if $\vecspace_0$ is convex, and this will allow us to design \E variables for such $\vecspace_0$.
Let $\cH_1 = \{P_{\theta^*_a,\theta^*_b} \}$ with 
$(\theta^*_a,\theta^*_b)$ in the interior of $[0,1]^2$, and let 
\begin{multline}\label{eq:klnormal}
\kl(\theta_a,\theta_b) := D(P_{\theta^*_a, \theta^*_b}(Y_a^{n_a}, Y_b^{n_b}) \| 
P_{\theta_a, \theta_b}(Y_a^{n_a}, Y_b^{n_b}) ) = \\ \sum_{y_a^{n_a} \in \{0,1\}^{n_a}, y_b^{n_b} \in \{0,1\}^{n_b}} 
p_{\theta^*_a}(y_a^{n_a})
p_{\theta^*_b} (y_b^{n_b}) \log \frac{p_{\theta^*_a}(y_a^{n_a})
p_{\theta^*_b} (y_b^{n_b}) }{p_{\theta_a}(y_a^{n_a})
p_{\theta_b} (y_b^{n_b}) } 
    \end{multline}
 stand for the KL divergence between $P_{\theta^*_a,\theta^*_b}$ and $P_{\theta_a, \theta_b}$ restricted to a single block (note that in the previous subsection, KL divergence was defined for a single outcome $Y$). 
The following result builds on Theorem~\ref{simpleSproportionsgenH0}:
\begin{theorem}\label{thm:convex}
$\min_{(\theta_a,\theta_b) \in \vecspace_0} \kl(\theta_a,\theta_b)$ is uniquely achieved by some  $(\theta^{\circ}_a,\theta^{\circ}_b)$.
If $(\theta^*_a,\theta^*_b) \in \vecspace_0$, then $(\theta^{\circ}_a,\theta^{\circ}_b)= (\theta^*_a,\theta^*_b)$. Otherwise,
$(\theta^{\circ}_a,\theta^{\circ}_b)$ lies on the boundary of $\vecspace_0$, but not on the boundary of $[0,1]^2$. The \E-variable (\ref{eq:simpleevargenH0}) is given by the distribution $W$ that puts all its mass on $(\theta^{\circ}_a,\theta^{\circ}_b)$, i.e. 
\begin{equation}\label{eq:evargennul}
s(y^{n_a}_a, y^{n_b}_b; n_a, n_b, (\theta^*_a,\theta^*_b); \vecspace_0)
=  \frac{p_{\theta_{a}^*}( y^{n_a}_a) p_{\theta_b^*}(y^{n_b}_b)}{
p_{\theta^{\circ}_{a}}( y^{n_a}_a) p_{\theta^{\circ}_b}(y^{n_b}_b)}
\end{equation}
is an \E variable. Moreover, this is the $(\theta^*_a,\theta^*_b)$-GRO \E variable relative to $\vecspace_0$.
\end{theorem}
We can extend this \E variable to the case of a composite $\cH_1 = \{P_{\theta_a,\theta_b}: 
(\theta_a,\theta_b) \in \vecpar_1\}$ by {\em learning\/} the true $(\theta^*_a,\theta^*_b) \in \vecpar_1$ from the data \citep{turner2021safe}. We thus replace, for each $j =1,2, \ldots$, for the block $Y_{(j)} $ consisting of $n_a$ points $Y_{(j),a,1}, \ldots, Y_{(j),a,n_a}$ in group $a$ and $n_b$ points $Y_{(j),b,1}, \ldots, Y_{(j),b,n_b}$ in group $b$, the `true' $\theta^*_g$ for $g \in \{a,b\}$ by an estimate $\breve\theta_g \mid Y^{(j-1)}$ based on the previous $j-1$ data blocks. 
The  \E variable corresponding to $m$ blocks of data then becomes
\begin{align}\label{eq:woensdag}
S^{(m)}_{[n_a,n_b,W_1; \vecspace_0]} & 
= 
\prod_{j=1}^m  
\prod_{i=1}^{n_a} \frac{p_{\breve\theta_a|Y^{(j-1)}}(Y_{(j),a,i})}{
p_{\breve\theta_a^{\circ} |Y^{(j-1)}}(Y_{(j),a,i})}
\prod_{i=1}^{n_b} 
\frac{p_{\breve\theta_b|Y^{(j-1)}}(Y_{(j),b, i})}{
p_{\breve\theta^{\circ}_b |Y^{(j-1)}}(Y_{(j), b,i})
}
\end{align}
where, for $g \in \{a,b\}$,  $\breve\theta_g|Y^{(j-1)}$ 
can be an arbitrary estimator (function from $Y^{(j-1)}$ to $\theta_g$)
and 
$(\breve\theta^{\circ}_a \mid Y^{(j-1)}, \breve\theta^{\circ}_b \mid Y^{(j-1)})$ is defined to achieve 
$\min_{(\theta_a,\theta_b) \in \vecspace_0} 
D(P_{\breve\theta_a \mid Y^{(j-1)},\breve\theta_b \mid Y^{(j-1)}}(Y^{n_a}_a,Y^{n_b}_b) \| P_{\theta_a,\theta_b}(Y^{n_a}_a,Y^{n_b}_b))$. 
No matter what estimator we choose, (\ref{eq:woensdag}) gives us an \E variable. In Section~\ref{sec:anytime}, as in \citep{turner2021safe}, we implement this estimator by fixing a prior $W$  and using the Bayes posterior mean,  $\breve\theta_g | Y^{(j-1)} := {\bf E}_{\theta_g \sim W \mid Y^{(j-1)}}[\theta_g]$.
Let us now illustrate Theorem~\ref{thm:convex} for two choices of $\vecspace_0$.
\paragraph{$\vecspace_0$ with linear boundary}
First, we let $\vecspace_0(s,c)$, for $s\in \reals, c \in \reals$, stand for any  straight line  through \\ $[0,1]^2$ : $\vecspace_0(s,c) := \{ (\sp) \in [0,1]^2: \theta_b = s + c \theta_a \}$.
This can be extended to $\vecspace_0({\leq} s,c) := \bigcup_{s' \leq s} \vecspace_0(s',c)$ and similarly to $\vecspace_0({\geq} s,c) := \bigcup_{s'\geq s} \vecspace_0(s',c)$. 
For example, we could take  $\vecspace_0 = \vecspace_0(s,c)$ to be the solid line in Figure \ref{fig:parameterSpaceExamples}(a) (which would correspond to $s = 0.1, c=1$), or the whole area underneath the line ($\vecspace_0({\leq s},c)$) including the line itself, or the whole area above it including the line itself  ($\vecspace_0({\geq} s,c)$). 

Now consider a  $\vecspace_0(s,c)$ that has nonempty intersection with the interior of $[0,1]^2$ and that is separated from the point alternative $(\theta^*_a,\theta^*_b)$, i.e.  $\min_{(\theta_a,\theta_b) \in \vecspace_0} 
\kl(\theta_a,\theta_b) > 0$. Simple differentiation gives that the minimum is achieved  by the unique $(\theta_a^{\circ},\theta_b^{\circ}) \in \vecspace_0$ satisfying:
\begin{align}\label{eq:klminimizer}
     n_a \left(- \frac{\ta}{\theta_a^{\circ}}
+\frac{1-\ta}{1- \theta_a^{\circ}}
\right) + 
 n_b  \cdot c \cdot \left(- \frac{\tb}{\theta_b^{\circ}}
+\frac{1-\tb}{1- \theta_b^{\circ}}
\right) = 0,
\end{align}
which can now be plugged into the \E variable (\ref{eq:evargennul}) if the alternative is the simple alternative, or otherwise into its sequential form (\ref{eq:woensdag}). 
In the basic case in which $\vecspace_0 =  \{ (\sp) \in [0,1]^2: \theta_a= \theta_b$\}, the solution to (\ref{eq:klminimizer}) reduces to the familiar
$
\theta^{\circ}_a = \theta^{\circ}_b = ({n_a \ta + n_b \tb })/{n}
$ from \cite{turner2021safe}.

If 
$(\theta^*_a,\theta^*_b)$ lies above the line $\vecspace_0(s,c)$, then by Theorem~\ref{thm:convex}, $\min_{(\theta_a,\theta_b) \in \vecspace_0({\leq s},c)} 
\kl(\theta_a,\theta_b)$ must lie on  $\vecspace_0(s,c)$. Theorem~\ref{thm:convex} gives that it must  be achieved  by the $(\theta_a^{\circ},\theta_b^{\circ})$ satisfying (\ref{eq:klminimizer}). Similarly, if 
$(\theta^*_a,\theta^*_b)$ lies below the line $\vecspace_0(s,c)$, then $\min_{(\theta_a,\theta_b) \in \vecspace_0({\geq} s,c)}$  
$\kl(\theta_a,\theta_b)$ is again achieved  by the $(\theta_a^{\circ},\theta_b^{\circ})$ satisfying (\ref{eq:klminimizer}).
\paragraph{$\vecspace_0$ with log odds ratio boundary}
Similarly, we can consider $\vecspace_0(\delta), \vecspace_0({\leq \delta}), \vecspace_0({\geq} \delta)$ that correspond to a given log odds effect size $\delta$. That is, we now take
\begin{align*}
   \vecspace_0(\delta) & := \left\{(\theta_a,\theta_b) \in [0,1]^2: \log \frac{\theta_b(1- \theta_a)}{(1-\theta_b)\theta_a}
    = \delta\right \} \\
    \vecspace_0({\leq} \delta) & := \left\{(\theta_a,\theta_b) \in [0,1]^2: \log \frac{\theta_b(1- \theta_a)}{(1-\theta_b)\theta_a} \leq  \delta \right\} \\
    \vecspace_0({\geq} \delta) & := \left\{(\theta_a,\theta_b) \in [0,1]^2: \log \frac{\theta_b(1- \theta_a)}{(1-\theta_b)\theta_a} \geq  \delta\right\}. 
\end{align*}
For example, we could now take  $\vecspace_0 = \vecspace_0({\leq} \delta)$ to be the area under the curve (including the curve boundary itself) in Figure~\ref{fig:parameterSpaceExamples}(b), which would correspond to $\delta=2$.  
Now let $\delta$ and point alternative $(\theta^*_a,\theta^*_b)$
be such that $\delta >0$ and  $\vecspace_0({\leq} \delta)$ is separated from $(\theta^*_a,\theta^*_b)$, i.e.  $\min_{(\theta_a,\theta_b) \in \vecspace_0({\leq } \delta)} 
\kl(\theta_a,\theta_b) > 0$. Let $(\theta^{\circ}_a,\theta^{\circ}_b) :=
\arg \min_{(\theta_a,\theta_b) \in \vecspace_0(\delta)} 
\kl(\theta_a,\theta_b)$. As Figure~\ref{fig:parameterSpaceExamples}(b) suggests, $\vecspace_0({\leq} \delta)$ is convex. Theorem~\ref{thm:convex} now tells us that  $\min_{(\theta_a,\theta_b) \in  \vecspace_0({\leq} \delta)} 
\kl(\theta_a,\theta_b)$ is achieved by $(\theta^{\circ}_a,\theta^{\circ}_b)$.
Plugging these into (\ref{eq:evargennul}) thus gives us an \E variable.  $(\theta^{\circ}_a,\theta^{\circ}_b)$ can easily be determined numerically. Similarly, if $\delta < 0$, $\vecspace_0({\geq} \delta)$ is convex and closed and if $(\theta^*_a,\theta^*_b)$ is separated from $\vecspace_0({\geq} \delta)$, the  $(\theta^{\circ}_a,\theta^{\circ}_b)$ minimizing KL on $\vecspace_0(\delta)$ gives an \E variable relative to $\vecspace_0({\geq} \delta)$.
\section{Anytime-Valid Confidence for the $2 \times 2$ case}\label{sec:anytime}
We will now use the \E variables defined above to construct anytime-valid confidence sequences. Let $\delta= \delta(\theta_a,\theta_b)$ be a notion of effect size such as the log odds ratio (see above) or absolute risk $\theta_b-\theta_a$ or relative risk $\theta_b/\theta_a$.
A $(1-\alpha)$-{\em anytime-valid (AV) confidence sequence\/} \citep{darling1967confidence,howard2018uniform} is a sequence of random (i.e. determined by data) subsets $\text{\sc CS}_{\alpha,(1)},\text{\sc CS}_{\alpha,(2)},\ldots$ of $\Gamma$, with $\text{\sc CS}_{\alpha,(m)}$ being a function of the first $m$ data blocks $Y^{(m)}$,  such that for all $(\theta_a,\theta_b) \in [0,1]^2$,
$$
P_{\theta_a,\theta_b}\left( 
\exists m\in \naturals: \delta(\theta_a,\theta_b) \not \in \text{\sc CS}_{\alpha,(m)} \right)\leq \alpha. 
$$
We first consider the case in which  for all values $\gamma \in \Gamma$ that $\delta$ can take,
$\vecspace_0(\gamma) := \{ (\theta_a,\theta_b) \in [0,1]^2: \delta(\theta_a,\theta_b) = \gamma \}$ is a convex set, as it will be for absolute and relative risk. Fix a prior $W_1$ on $[0,1]^2$. Based on (\ref{eq:woensdag}) we can make an {\em exact\/} (nonasymptotic) AV  
confidence sequence 
\begin{equation}
    \label{eq:avci}
\text{\sc CS}_{\alpha,(m)} = \left\{ \delta : 
S^{(m)}_{[n_a,n_b,W_1;\vecspace_0(\delta)]} \leq \frac{1}{\alpha} \right\}
\end{equation}
where $S^{(m)}_{[n_a,n_b,W_1;\vecspace_0(\delta)]}$
is defined as in (\ref{eq:woensdag}) and is a valid \E variable by Theorem~\ref{thm:convex}. To see that $(\text{\sc CS}_{\alpha,(m)})_{m \in \naturals}$ really is an AV confidence sequence, note that, by definition of the $\text{\sc CS}_{\alpha,(m)}$, we have\\ $
 P_{\theta_a,\theta_b}\left( \exists m\in \naturals: \delta(\theta_a,\theta_b) \not \in \text{\sc CS}_{\alpha,(m)} \right)
$ is given by 
\begin{align*}
P_{\theta_a,\theta_b}\left( \exists m\in \naturals:
S^{(m)}_{[n_a,n_b,W_1;\vecspace_0(\delta)]} \geq \frac{1}{\alpha} \right) \leq \alpha, 
\end{align*}
by Ville's inequality \citep{grunwald2019safe,turner2021safe}. Here the $\text{\sc CS}_{\alpha,(m)}$ are not necessarily intervals, but, potentially loosing some information, we can make a  AV confidence sequence consisting of intervals by defining $\text{\sc CI}_{\alpha,(m)}$ to be the smallest interval containing $\text{CS}_{\alpha,(m)}$. We can also turn any confidence sequences $(\text{\sc CS}_{\alpha,(m)})_{m \in \naturals}$ 
into an alternative AV confidence sequence with sets $\text{\sc CS}'_{\alpha,(m)}$ that are always a subset of $\text{\sc CS}_{\alpha,(m)}$ by taking the {\em running intersection\/}
$$
\text{\sc CS}'_{\alpha,(m)} := \bigcap_{j =1.. m }\text{\sc CS}_{\alpha,(j)}.
$$
In this form, the confidence sequences $\text{\sc CS}'_{\alpha,(m)}$ can be interpreted as {\em the set of $\delta$'s that have not yet been rejected\/} in a setting in which, for each null hypothesis $\vecspace_0(\delta)$ we stop and reject as soon as
the corresponding \E variable exceeds $1/\alpha$. The running intersection can also be applied to the intervals $(\text{\sc CI}_{\alpha,(m)})_{m \in \naturals}$.

To simplify calculations, it is useful to take $W_1$ a prior under which $\theta_a$ and $\theta_b$ have independent beta distributions with parameters $\alpha_a,\beta_a, \alpha_b,\beta_b$. We can, if we want, infuse some prior knowledge or hopes by setting these parameters to certain values --- our confidence sequences will be valid irrespective of our choice \citep{howard2018uniform}. 
In case no such knowledge can be formulated (as in the simulations below), we advocate the prior, which, among all priors of the simple form asymptotically achieves the REGROW criterion (a criterion related to minimax log-loss regret, see \citep{grunwald2019safe}), i.e for the case $n_a=n_b=1$ we set $W_1$ to an independent beta prior on $\theta_a$ and $\theta_b$ with  $\gamma= 0.18$ as was empirically found to be the `best' value \citep{turner2021safe}. 

\paragraph{Log Odds Ratio Effect Size} The situation is slightly trickier if we take the log odds ratio as effect size, for $\vecspace_0(\delta)$ is then not convex. Without convexity, Theorem~\ref{thm:convex} cannot be used and hence the validity of AV confidence sequences as constructed above breaks down. 
We can get nonasymptotic anytime-valid confidence sequences after all as follows. First, we consider a one-sided AV confidence sequence for the submodel of positive effect sizes $\{(\theta_a,\theta_b): \delta(\theta_a,\theta_b) \geq 0\}$, defining 
$$\text{\sc CS}^+_{\alpha,(m)}= \{ \delta \geq 0:  S^{(m)}_{[n_a,n_b,W_1;\vecspace_0(\leq \delta)]} \leq \alpha^{-1},
\}$$
where we note that $\vecspace_0(\leq \delta)$ is convex (since $\delta \geq 0$) and also contains $(\theta_a,\theta_b)$ with $\delta(\theta_a,\theta_b) < 0$. This confidence sequence can give a lower bound on $\delta$. Analogously, 
we consider a one-sided AV confidence sequence for the submodel $\{(\theta_a,\theta_b): \delta(\theta_a,\theta_b) \leq 0\}$, defining 
$$\text{\sc CS}^-_{\alpha,(m)}= \{ \delta \leq 0:  S^{(m)}_{[n_a,n_b,W_1;\vecspace_0(\geq \delta)]} \leq \alpha^{-1}
\},$$
and derive an upper bound on $\delta$.
By Theorem~\ref{thm:convex}, both sequences $(\text{\sc CS}^+_{\alpha,(m)})_{m=1,2,\ldots}$ and $(\text{\sc CS}^-_{\alpha,(m)})_{m=1,2,\ldots}$ are AV confidence sequences for the submodels with $\delta \geq 0$ and $\delta \leq 0$ respectively. Defining  $\text{\sc CS}_{\alpha,(m)} = \text{\sc CS}^+_{\alpha,(m)} \cup \text{\sc CS}^-_{\alpha,(m)}$, we find, for $(\theta_a,\theta_b)$ with $\delta(\theta_a,\theta_b) > 0$, 
$$
P_{\theta_a,\theta_b}\left( 
\exists m\in \naturals: \delta(\theta_a,\theta_b) \not \in \text{\sc CS}_{\alpha,(m)} \right)
= P_{\theta_a,\theta_b}\left( 
\exists m\in \naturals: \delta(\theta_a,\theta_b) \not \in \text{\sc CS}^+_{\alpha,(m)} \right)
\leq \alpha, 
$$
and analogously  for $ (\theta_a,\theta_b)$ with $\delta(\theta_a,\theta_b) < 0$.
We have thus arrived at a confidence sequence that works for all $\delta$, positive or negative.  
\subsection{Simulations}
\label{sec:simulations}
\begin{figure}[ht]
     \centering
     \begin{subfigure}[b]{9cm}
    \centering
    \includegraphics[width =\textwidth]{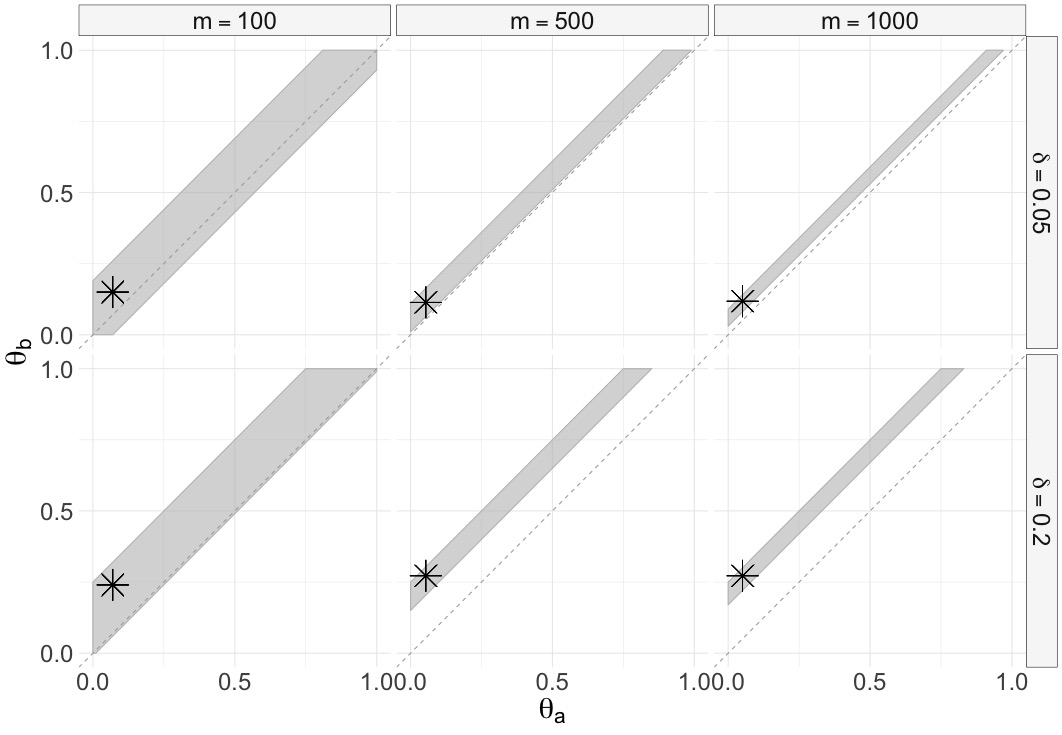}
    \caption{Risk difference}
     \end{subfigure}
     \hspace{1cm}
     \begin{subfigure}[b]{9cm}
    \centering
    \includegraphics[width =\textwidth]{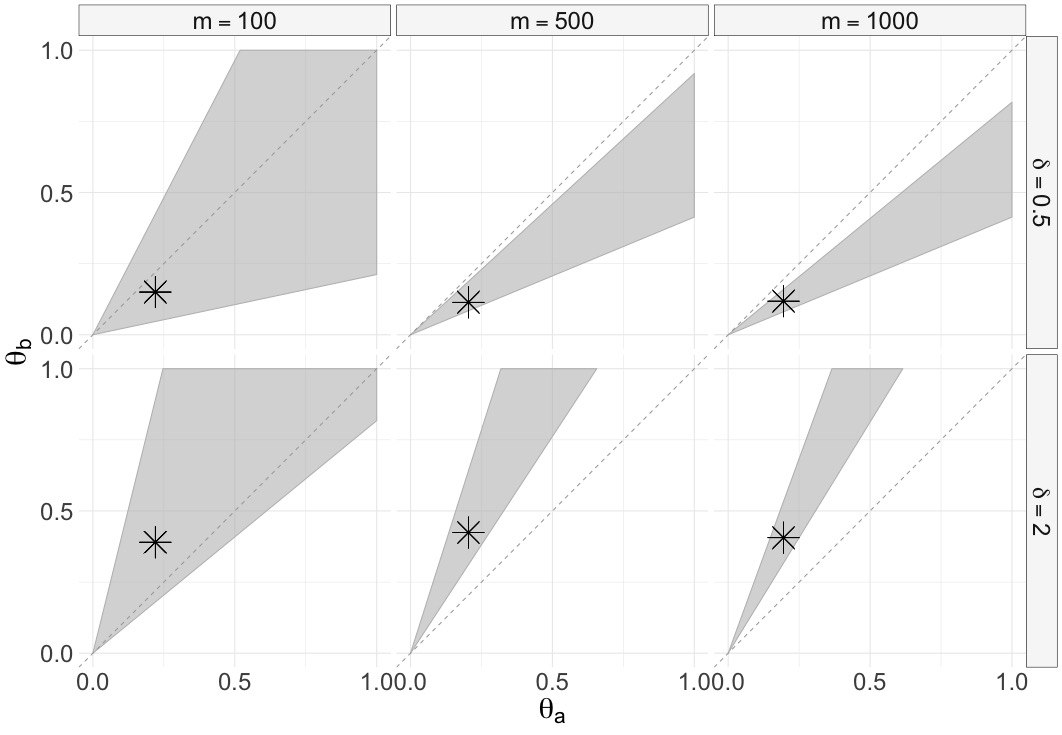}
    \caption{Relative risk}
     \end{subfigure}
    \caption{Depiction of parameter space with running intersection of confidence sequence for data generated under various effect sizes, at different time points $m$ in a data stream. The asterisks indicate the maximum likelihood estimator at that time point. The significance threshold was set to $0.05$. The design was balanced, with data block sizes $n_a = 1$ and $n_b = 1$.}
    \label{fig:beams}
\end{figure}

In this section some numerical examples of confidence sequences for the two types of effect sizes are given. All simulations were run with code available in our software package \citep{ly2022safestats}.
\paragraph{Risk difference} 
Risk difference is defined as the difference between success probabilities in the two streams: $\delta = \theta_b - \theta_a$. Figure~\ref{fig:beams} shows running intersections of confidence sequences with $\delta$ as the risk difference for simulations for various distributions and stream lengths. These sequences are constructed by testing null hypotheses based on $\vecspace_0(s,c)$, with $c = 1$ and $s = \delta$.  $\text{\sc CI}_{\alpha,(m)}$ for the risk difference on $\vecspace_0$ is an interval, corresponding to the `beam' of $(\theta_a,\theta_b) \in [0,1]^2$ bounded by the lines $\theta_b =\theta_a + \delta_{\text{\sc l}}$ and $\theta_b =\theta_a + \delta_{\text{\sc r}}$ with $\delta_{\text{\sc l}}>\delta_{\text{\sc r}}$ being values such that $S^{(m)}_{[n_a,n_b,W_1;\vecspace_0(\delta_{\text{\sc l}})]} =
S^{(m)}_{[n_a,n_b,W_1;\vecspace_0(\delta_{\text{\sc r}})]} = 1/\alpha
$. Figure \ref{fig:running} in the Appendix illustrates that the running intersection indeed improves the confidence sequence, albeit slightly.
\paragraph{Relative risk} Relative risk is defined as the ratio between the success probabilities in group $b$ and $a$: $\delta = \theta_b / \theta_a$. Hence, confidence sequences for this effect size measure can again be constructed using the linear boundary form $\vecspace_0(s,c)$ again, but now with $s = 0$ and $c = \delta$. Figure~\ref{fig:beams} shows running intersections of confidence sequences with $\delta$ as the relative risk.

\begin{figure}[ht]
     \centering
     \begin{subfigure}[b]{4cm}
         \centering
         \includegraphics[width=\textwidth]{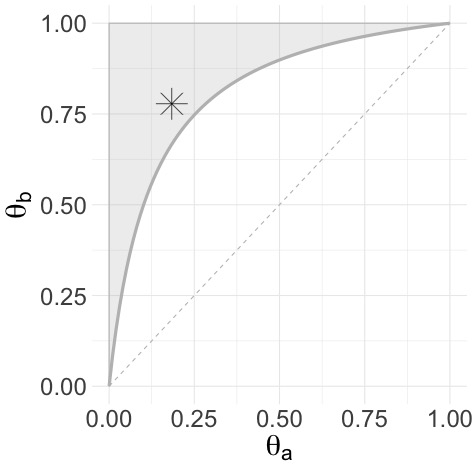}
         \caption{$CS^+$ at $n = 500$, true lOR $2.5$}
     \end{subfigure}
     \hspace{2cm}
     \begin{subfigure}[b]{4cm}
         \centering
         \includegraphics[width=\textwidth]{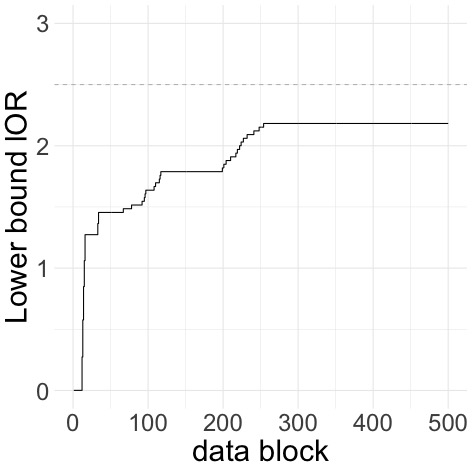}
         \caption{Running lower bound $CS^+$, true lOR $2.5$}
     \end{subfigure} \\
     \begin{subfigure}[b]{4cm}
         \centering
         \includegraphics[width=\textwidth]{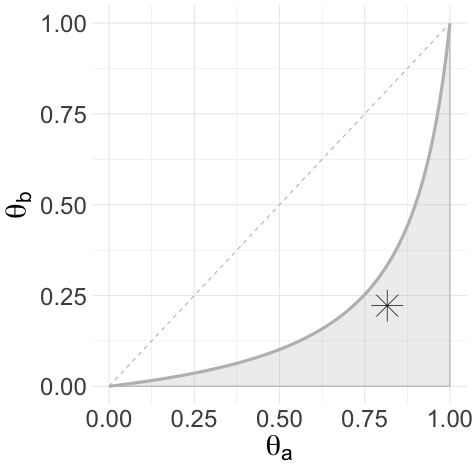}
         \caption{$CS^-$ at $n = 500$, true lOR $-2.5$}
     \end{subfigure}
     \hspace{2cm}
     \begin{subfigure}[b]{4cm}
         \centering
         \includegraphics[width=\textwidth]{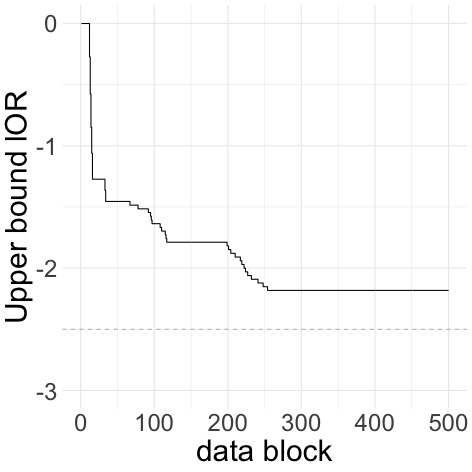}
         \caption{Running upper bound $CS^-$, true lOR $-2.5$}
     \end{subfigure}
    \caption{One-sided confidence sequences for odds ratios. 500 data blocks were generated under $P_{\theta_a, \theta_b}$ with $\theta_a = 0.2$ and log of the odds ratio (lOR) $2.5$ for figures a and b, and $\theta_a = 0.8$ and lOR $-2.5$ for figures c and d. The asterisks indicate the maximum likelihood estimator at $n = 500$. The significance threshold was set to $0.05$. The design was balanced, with data block sizes $n_a = 1$ and $n_b = 1$. Note that $CS^-$ is empty for (a) and (b) and $CS^+$ for (c) and (d) in these  confidence sequences.}
    \label{fig:odds}
\end{figure}

\paragraph{Log odds ratio boundary} If the maximum likelihood estimate based on $Y^{(m)}$ lies in the upper left corner as in Figure~\ref{fig:odds}(a), the confidence sets $\text{\sc CS}_{(m)}$ we get at time $m$ have a one-sided shape such as the shaded region, or the shaded region in Figure~\ref{fig:odds}(c), if the estimate lies in the lower right corner. Again, we can improve these confidence sequences by taking the running intersection; running intersections over time are illustrated in Figures \ref{fig:odds}(b) and \ref{fig:odds}(d).
\clearpage
\section{Conclusion}
We have shown how \E variables for data streams can be extended to general null hypotheses and non-asymptotic always-valid confidence sequences. We specifically implemented the confidence sequences for the $2 \times 2$ contingency tables setting; the resulting confidence sequences are efficiently computed and show quick convergence in simulations. For estimating risk differences or relative risk ratios between proportions in two groups, to our knowledge, such exact confidence sequences did not yet exist. For the log odds ratio we could also have used the sequential probability ratio (SPR) in Wald's SPR test \citep{wald1945sequential} test, which can be re-interpreted as a (product of) \E variables \citep{grunwald2019safe}. However, the SPR does not satisfy the GRO property making it sub-optimal (see also \citep{Adams20}); moreover, as should be clear from the development, our method for constructing confidence sequences can be implemented for any effect size notion with convex rejection sets $\vecspace_0(\leq \delta)$  and $\vecspace_0(\geq \delta)$, not just the log odds ratio. A main goal for future work is to use Theorem~\ref{thm:convex} to provide such sequences for sequential two-sample settings that go beyond the $2 \times 2$ table. 
\commentout{
Although the \E variables described in this paper would in theory approximate the GRO \E variable quickly when the parameters of the real data generating distribution were estimated quickly, the question remains if one could improve the convergence of these sequences for example through conditioning, for example on the sum of positive outcomes, although it is not directly clear if conditional \E variables for general null hypotheses would have such a simple analytical expression as the ones described in this paper \citep{Adams20}.
}
\section*{Acknowledgements}
Funding: this work is part of the Enabling Personalized Interventions (EPI) project, which is supported by the Dutch Research Council (NWO) in the Commit2 - Data –Data2Person program under contract 628.011.028. Declarations of interest: none.
\bibliographystyle{plainnat}
\bibliography{references}
\newpage
\begin{appendix}
\appendixpage
\renewcommand{\thesection}{\Alph{section}}
\renewcommand\thefigure{\thesection.\arabic{figure}} 
\section{Proofs}
\label{app:proofs}
Both proofs below  use Theorem 1 of \cite{grunwald2019safe} and a direct corollary (called Corollary 2 by \cite{grunwald2019safe}), which we re-state here, for convenience, combined as a single statement. Recall that we use notation $P_W := \int P_{\vecpar} d W(\vecpar)$.
\paragraph{Theorem (Theorem 1 of \cite{grunwald2019safe})} 
Let $Y$ be a random variable taking values in a set ${\cal Y}$. Suppose $Q$ is a probability distribution for $Y$ with density $q$ that is strictly positive on all of ${\cal Y}$ and let $\cH_0 = \{ P_{\vecpar}: \vecpar \in \vecspace_0 \}$  be a set of distributions for $Y$ where each $P_{\vecpar}$ has density $p_{\vecpar}$. Let ${\cal W}_0$ be the set of all distributions on $\vecspace_0$. Assume  $\inf_{W_0 \in {\cal W}_0(\vecspace_0)} D(Q  \| P_{W_0}) < \infty$.  Then (a) 
there exists a (potentially sub-) distribution $P^*_0$ with density $p^*_0$ such that 
\begin{equation*}
S^{*} := \frac{q(Y)}{p^*_0(Y)}
\end{equation*} 
is an \E variable ($p^*_0$ is called the {\em Reverse Information Projection (RIPr) of $q$ onto $\{ p_{W}: W \in {\cal W}_0 \}$}).
Moreover, (b), $S^*$ satisfies 
\begin{align}\label{eq:firstgro}
  \sup_{S \in {\cal E}(\vecspace_0)}  {\bf E}_{Y \sim Q}[\log S]
  =  
  {\bf E}_{Y \sim Q}[\log S^*] =
  \inf_{W_0 \in {\cal W}_0(\vecspace_0)}  D(Q \| P_{W_0})
  = D(Q \| P^*_0).  
  \end{align} 
(where ${\cal E}(\vecspace_0)$ is the set of all \E variables relative to null hypothesis $\cH_0$) and $S^*$ is thus the $Q$-GRO \E variable for $Y$. 
If the minimum is achieved by some ${W}^*_0$, i.e.\ $D(Q \| P^*_0) = D(Q \|
  P_{W^*_0})$, then $P^*_0 = P_{W^*_0}$. 
Moreover, (c), if there exists an \E variable $S$ of the form $q(Y)/p_{W_0}(Y)$ for some $W_0 \in {\cal W}_0$ then $W_0$ must achieve the infimum in (\ref{eq:firstgro}) and $S$ must be essentially equal to $S^*$ in the sense that for  all $P \in \cH_0 \cup \{Q \}$, $P(S^* = q(Y)/p_{W_0}(Y)) = 1$. Similarly (d), if there exists a $W^*_0 \in {\cal W}_0$ that achieves the infimum in (\ref{eq:firstgro}) then $S = q(Y)/p_{W^*_0}(Y)$ is an \E variable and $S$ is again essentially equal to $S^*$. 
\paragraph{Proof of Theorem~\ref{simpleSproportionsgenH0}}
{\em Part 1\/} The real idea behind the proof is the formulation of the modified testing problem in which only a single outcome per block is observed. This we already did in the main text. Linking the two is simply the last, very simple step, with analogies to the proof of Part 1 of Theorem 1 in \cite{turner2021safe}.

Let $n_a, n_b \in \naturals, n := n_a+n_b$ and let $u,v \in \reals^+$. Suppose that $n_a u + n_b v \leq n$.
Then $u^{n_a} v^{n_b} \leq 1$, which follows immediately from applying Young's inequality to $u^{n_a/n},v^{n_b/n}$ but can also be derived directly by writing $v$ as function of $u$ and differentiating $\log (u^{n_a} v^{n_b}) $ to $u$.

Further, by independence, for $(\theta_a,\theta_b) \in \vecspace_0$,
\begin{align}
  &  {\bf E}_{Y_a^{n_a} \sim  P_{\theta_a}, Y_b^{n_b} \sim P_{\theta_b}}\left[
  s'(Y^{n_a}_a, Y^{n_b}_b)
  \right] 
  = \nonumber \\ & {\bf E}_{Y_a^{n_a} \sim  P_{\theta_a}}\left[
  \frac{p_{\theta_{a}^*}( Y^{n_a}_a) }{
p^{\circ}(Y^{n_a}_a |a)}
  \right]  \cdot {\bf E}_{Y_b^{n_b} \sim P_{\theta_b}}\left[
   \frac{ p_{\theta_b^*}(Y^{n_b}_b)}{
p^{\circ}(Y^{n_b}_b |b)}
  \right] \nonumber = \\ \nonumber
  &  \left({\bf E}_{Y\sim  P_{\theta_a}}\left[
  \frac{p_{\theta_{a}^*}( Y) }{
p^{\circ}(Y |a)}
  \right] \right)^{n_a} \cdot 
  \left({\bf E}_{Y \sim P_{\theta_b}}\left[
   \frac{ p_{\theta_b^*}(Y)}{
p^{\circ}(Y |b)}
  \right] \right)^{n_b} = \\ 
  &  \left({\bf E}_{Y \sim  P'_{\theta}|a}\left[
  \frac{p'_{\theta^*}( Y|a) }{
p^{\circ}(Y |a)}
  \right] \right)^{n_a} \cdot 
  \left({\bf E}_{Y \sim P'_{\theta|b}}\left[
   \frac{ p'_{\theta^*}(Y|b)}{
p^{\circ}(Y |b)}
  \right] \right)^{n_b}. 
\end{align}

Combining the two facts stated above, (\ref{eq:maandag}) implies that the latter quantity is bounded by 1. 

{\em Part 2\/} By lower-semicontinuity of the KL divergence in its second argument (Posner's theorem, used as in \cite{grunwald2019safe}) the infimum in (\ref{eq:klmina}) is achieved by some prior distribution $W^{\circ}$ so that  by Theorem~1 of \cite{grunwald2019safe} (part (b) in the formulation above), $p^{\circ}(\cdot \mid \cdot) = p'_{W^{\circ}}(\cdot \mid \cdot)$ and hence also $P^{\circ}(G,Y) = P'_{W^{\circ}}(G,Y)$. By convexity of $\cH'_0$ and finiteness of the support of $P'_{\vecpar}(G,Y)$, there must be some $\vecpar$ such that $P'_{W^{\circ}}(G,Y) = P_{\vecpar}(G,Y)$ and hence also  $p'_{W^{\circ}}(\cdot \mid \cdot)= p'_{\vecpar}(\cdot \mid \cdot)$, which shows (a). This means that we have now created an \E variable for the original problem which can be written as 
$p_{\theta^*_a,\theta^*_b}/p_{W_0}$ with $p_{W_0}$ a prior distribution on $\vecpar_0$ (namely, the one that puts mass $1$ on $\vecpar$). 
(b) is then an immediate consequence of Theorem~1 of \cite{grunwald2019safe} (part (c) in the formulation above). 
(note that we {\em cannot\/} draw this conclusion if $\cH'_0$ is not convex; for then the distribution $p'_{W^{\circ}}$ may not correspond to the distribution $p_{W^{\circ}}$ in the original problem --- this correspondence is only guaranteed if $p'_{W^{\circ}}$ coincides with some $p'_{\vec{\theta}}$. 
\paragraph{Proof of Theorem~\ref{thm:convex}}
Recall that we assume that $\vecspace_0$ is convex and compact. We set $\kl'(\theta_a,\theta_b) := D(P'_{\theta^*_a,\theta^*_b} \| P'_{\theta_a,\theta_b} )$ where $D$ is the KL divergence  as in (\ref{eq:klminb}), i.e. for the modified setting in which $P'_{\theta_a,\theta_b}$ is a distribution on a single outcome, as discussed before Theorem~\ref{simpleSproportionsgenH0}. For the $2 \times 2$ model this KL divergence can be written explicitly as 
\begin{equation}
    \label{eq:klmodified}
 D(P'_{\theta^*_a,\theta^*_b} \| P'_{\theta_a,\theta_b} ) = 
 \frac{n_a}{n} 
 \sum_{y_a \in \{0,1\}}
p_{\theta^*_a}(y_a) 
\log \frac{p_{\theta^*_a}(y_a) }{p_{\theta_a}(y_a)
 } 
 + \frac{n_b}{n}  \sum{y_b  \in \{0,1\} }
p_{\theta^*_b} (y_b)  \log \frac{
p_{\theta^*_b} (y_b) }{
p_{\theta_b} (y_b) } 
\end{equation}
From (\ref{eq:klnormal}) we now see that $n \kl'(\theta_a,\theta_b) = \kl(\theta_a,\theta_b)$. We will prove the theorem with $\kl$ replaced by $\kl'$ and $\cH_0$ by $\cH'_0$; since the two KL's agree up to a constant factor of $n$, all results transfer to the $\kl$ mentioned in the theorem statement. 

Since $\vecspace_0$ is compact in the Euclidean topology and all distributions in $\cH'_0$ can be represented as 2-dimensional vectors, i.e. they have common and finite support, we must have that $\cH_0$ is compact in the weak topology so we can use the lower-semicontinuity of KL divergence in its second argument (Posner's theorem) as in \citep{grunwald2019safe} to give us that the minimum KL divergence $\min \kl'(\theta_a,\theta_b)$ is  achieved by some $(\theta^{\circ}_a,\theta^{\circ}_b)$. Since KL divergence is strictly convex in its second argument 
and $\cH'_0$ is convex (this is the place where we need to use $\kl'$ rather than $\kl$: $\cH_0$ may  {\em not\/} be convex!), the minimum must be achieved uniquely. 
Since KL divergence $\kl'(\theta_a,\theta_b)$ is nonnegative and $0$ only if $(\theta_a,\theta_b) = (\theta^*_a,\theta^*_b)$, it follows that $(\theta^{\circ}_a,\theta^{\circ}_b) = (\theta^*_a,\theta^*_b)$ if $\min \kl(\theta_a,\theta_b)= 0$. Otherwise, 
since we assume $(\theta^*_a,\theta^*_b)$ to be  in the interior of $[0,1]^2$, $ \kl(\theta_a,\theta_b) = \infty$ iff $(\theta_a,\theta_b)$ lies on the boundary of $[0,1]^2$. Thus, $(\theta^{\circ}_a,\theta^{\circ}_b)$ must lie in the interior of $[0,1]^2$ as well. $(\theta^{\circ}_a,\theta^{\circ}_b)$  cannot lie in the interior of $\vecspace_0$ though: for any 
point $(\theta_a,\theta_b)$ in the interior of $\vecspace_0$ we can draw a line segment between this point and $(\theta^*_a,\theta^*_b)$. Differentiation along that line gives that $\kl'(\theta_a,\theta_b)$ monotonically decreases as we move towards $(\theta^*_a,\theta^*_b)$, so the  minimum within the closed set $\vecspace_0$ must lie on its boundary. 

It remains to show that (\ref{eq:evargennul}) is the $(\theta^*_a,\theta^*_b)$-GRO \E variable relative to $\cH_0$. To see this, note that, by convexity of $\cH'_0$, from Theorem~\ref{simpleSproportionsgenH0}, we must have that the GRO \E variable for this original problem is of the form 
$$
\frac{p_{\theta_{a}^*}( y^{n_a}_a) p_{\theta_b^*}(y^{n_b}_b)}{
p_{\theta^{+}_a}(y^{n_a}_a )
p_{\theta^{+}_b}(y^{n_b}_b )}
$$
for some $(\theta^{+}_a,\theta^{+}_b)$.
The result then follows again by Theorem~1 of \cite{grunwald2019safe} (part (c) in the formulation above): this shows that the distribution $W_0$ that puts mass 1 on $(\theta^{+}_a,\theta^{+}_b)$ minimizes, among all distributions $W$ on $\vecspace_0$, $D(P_{\theta^*_a,\theta^*_b} \| P_{W})$.
Since the set of such distributions includes all distributions that put mass $1$ on {\em some\/} $(\theta_a,\theta_b) \in \vecspace_0$, we must have that $(\theta^{+}_a,\theta^{+}_b)= (\theta^{\circ}_a,\theta^{\circ}_b)$.
\newpage
\section{Extended simulation results}
\renewcommand\thefigure{\thesection.\arabic{figure}}
\setcounter{figure}{0} 
\begin{figure}[ht]
    \centering
    \includegraphics[width = 6cm]{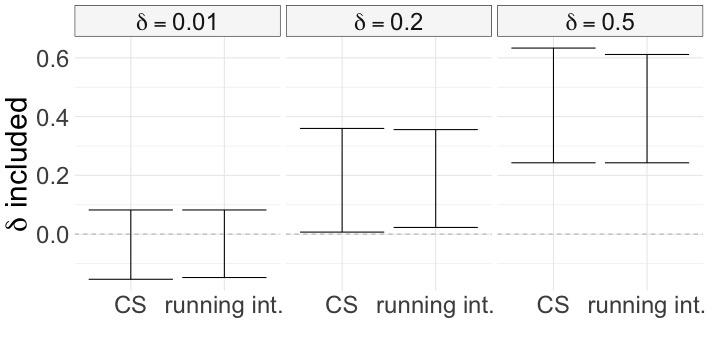}
    \caption{Confidence sequence with and without running intersection, for data generated under $P_{\theta_a, \theta_a + \delta}$ with $\theta_a = 0.05$, for a data stream of length $100$. The significance threshold was set to $0.05$. The design was balanced, with data block sizes $n_a = 1$ and $n_b = 1$.}
    \label{fig:running}
\end{figure}
\end{appendix}

\end{document}